# IRGAN: A Minimax Game for Unifying Generative and Discriminative Information Retrieval Models

Jun Wang
University College London
j.wang@cs.ucl.ac.uk

Lantao Yu, Weinan Zhang*
Shanghai Jiao Tong University
wnzhang@sjtu.edu.cn

Yu Gong, Yinghui Xu
Alibaba Group
renji.xyh@taobao.com

Benyou Wang, Peng Zhang
Tianjin University
pzhang@tju.edu.cn

Dell Zhang
Birkbeck, University of London
dell.z@ieee.org

## ABSTRACT

This paper provides a unified account of two schools of thinking in information retrieval modelling: the generative retrieval focusing on predicting relevant documents given a query, and the discriminative retrieval focusing on predicting relevancy given a query-document pair. We propose a game theoretical *minimax game* to iteratively optimise both models. On one hand, the discriminative model, aiming to mine signals from labelled and unlabelled data, provides guidance to train the generative model towards fitting the underlying relevance distribution over documents given the query. On the other hand, the generative model, acting as an attacker to the current discriminative model, generates difficult examples for the discriminative model in an adversarial way by minimising its discrimination objective. With the competition between these two models, we show that the unified framework takes advantage of both schools of thinking: (i) the generative model learns to fit the relevance distribution over documents via the signals from the discriminative model, and (ii) the discriminative model is able to exploit the unlabelled data selected by the generative model to achieve a better estimation for document ranking. Our experimental results have demonstrated significant performance gains as much as 23.96% on Precision@5 and 15.50% on MAP over strong baselines in a variety of applications including web search, item recommendation, and question answering.

## 1 INTRODUCTION

A typical formulation of information retrieval (IR) is to provide a (rank) list of documents given a query. It has a wide range of applications from text retrieval [1] and web search [3, 19] to recommender systems [21, 34], question answering [9], and personalised advertising [27], to name just a few. There are, arguably, two major schools of thinking when coming to IR theory and modelling [1].

The classic school of thinking is to assume that there is an underlying stochastic **generative** process between documents and information needs (clued by a query) [22]. In text retrieval, the classic relevance model of information retrieval is focused on describing how a (relevant) document is generated from a given information need: $q \to d$, where $q$ is the query (e.g., keywords, user profiles, questions, depending on the specific IR application), $d$ is its corresponding document (e.g., textual documents, information items, answers), and the arrow indicates the direction of generation. Notable examples include Robertson and Sparck Jones's Binary Independence Model, of which each word token is independently generated to form a relevant document [35]. Statistical language models of text retrieval consider a reverse generative process from a document to a query: $d \to q$, typically generating query terms from a document (i.e., the query likelihood function) [32, 48]. In the related work of word embedding, word tokens are generated from their context words [28]. In the application of recommender systems, we also see that a recommended target item (in the original document identifier space) can be generated/selected from known context items [2].

The modern school of thinking in IR recognises the strength of machine learning and shifts to a **discriminative** (classification) solution learned from labelled relevant judgements or their proxies such as clicks or ratings. It considers documents and queries jointly as features and predicts their relevancy or rank order labels from a large amount of training data: $q + d \to r$, where $r$ denotes relevance and symbol + denotes the combining of features. A significant development in web search is learning to rank (LTR) [3, 19], a family of machine learning techniques where the training objective is to provide the right ranking order of a list of documents (or items) for a given query (or context) [24]. Three major paradigms of learning to rank are pointwise, pairwise, and listwise. Pointwise methods learn to approximate the relevance estimation of each document to the human rating [23, 31]. Pairwise methods aim to identify the more-relevant document from any document pair [3]. Listwise methods learn to optimise the (smoothed) loss function defined over the whole ranking list for each query [4, 6]. Besides, a recent advance in recommender systems is matrix factorisation, where the interactive patterns of user features and item features are exploited via vector inner product to make the prediction of relevancy [21, 34, 46].

While the generative models of information retrieval are theoretically sound and very successful in modelling features (e.g., text statistics, distribution over document identifier space), they suffer from the difficulty in leveraging relevancy signals from other channels such as links, clicks etc., which are largely observable in Internet-based applications. While the discriminative models of information retrieval such as learning to rank are able to learn a retrieval ranking function implicitly from a large amount of labelled/unlabelled data, they currently lack a principled way of obtaining useful features or gathering helpful signals from the massive unlabelled data available, in particular, from text statistics (derived from both documents and queries) or the distribution of relevant documents in the collection.

In this paper, we consider the generative and discriminative retrieval models as two sides of the same coin. Inspired by Generative Adversarial Nets (GANs) in machine learning [13], we propose a game theoretical *minimax game* to combine the above mentioned two schools of thinking. Specifically, we define a common retrieval function (e.g., discrimination-based objective function) for both models. On one hand, the discriminative model $p_\phi(r|q,d)$ aims to maximise the objective function by learning from labelled data. It naturally provides alternative guidance to the generative retrieval model beyond traditional log-likelihood. On the other hand, the generative retrieval model $p_\theta(d|q,r)$ acts as a challenger who constantly pushes the discriminator to its limit. Iteratively it provides the most difficult cases for the discriminator to retrain itself by adversarially minimising the objective function. In such a way, the two types of IR models act as two players in a minimax game, and each of them strikes to improve itself to 'beat' the other one at every round of this competition. Note that our minimax game based approach is fundamentally different from the existing game-theoretic IR methods [26, 47], in the sense that the existing approaches generally try to model the interaction between user and system, whereas our approach aims to unify generative and discriminative IR models.

---
*The corresponding authors: J. Wang and W. Zhang.

Empirically, we have realised the proposed minimax retrieval framework in three typical IR applications: web search, item recommendation, and question answering. In our experiments, we found that the minimax game arrives at different equilibria and thus different effects of unification in different settings. With the pointwise adversarial training, the generative retrieval model can be significantly boosted by the training rewards from the discriminative retrieval model. The resulting model outperforms several strong baselines by 22.56% in web search and 14.38% in item recommendation on Precesion@5. We also found that with new pairwise adversarial training, the discriminative retrieval model is largely boosted by examples selected by the generative retrieval model and outperforms the compared strong algorithms by 23.96% on Precision@5 in web search and 2.38% on Precision@1 in question answering.

## 2 IRGAN FORMULATION

In this section, we take the inspiration from GANs and build a unified framework for fusing generative and discriminative IR in an adversarial setting; we call it IRGAN, and its application to concrete IR problems will be given in the next section.

### 2.1 A Minimax Retrieval Framework

Without loss of generality, let us consider the following information retrieval problem. We have a set of queries $\{q_1, ..., q_N\}$ and a set of documents $\{d_1, ..., d_M\}$. In a general setting, a query is any specific form of the user's information need such as search keywords, a user profile, or a question, while documents could be textual documents, information items, or answers, depending on the specific retrieval task. For a given query $q_n$, we have a set of relevant documents labelled, the size of which is much smaller than the total number of documents $M$.

The underlying *true* relevance distribution can be expressed as conditional probability $p_{\text{true}}(d|q,r)$, which depicts the (user's) relevance preference distribution over the candidate documents with respect to her submitted query. Given a set of samples from $p_{\text{true}}(d|q,r)$ observed as the training data, we can try to construct two types of IR models:

**Generative retrieval model** $p_\theta(d|q,r)$, which tries to generate (or select) relevant documents, from the candidate pool for the given query $q$, as specified later in Eq. (8); in other words, its goal is to approximate the true relevance distribution over documents $p_{\text{true}}(d|q,r)$ as much as possible.

**Discriminative retrieval model** $f_\phi(q,d)$, which, in contrary, tries to discriminate well-matched query-document tuples $(q,d)$ from ill-matched ones, where the goodness of matching given by $f_\phi(q,d)$ depends on the relevance of $d$ to $q$; in other words, its goal is to distinguish between relevant documents and non-relevant documents for the query $q$ as accurately as possible. It is in fact simply a binary classifier, and we could use 1 as the class label for the query-document tuples that truly match (positive examples) while 0 as the class label for those that do not really match (negative examples).

*2.1.1 Overall Objective.* Thus, inspired by the idea of GAN, we aim to unify these two different types of IR models by letting them play a minimax game: the generative retrieval model would try to generate (or select) relevant documents that look like the ground-truth relevant documents and therefore could fool the discriminative retrieval model, whereas the discriminative retrieval model would try to draw a clear distinction between the ground-truth relevant documents and the generated ones made by its opponent generative retrieval model. Formally, we have:

$$J^{G^*, D^*} = \min_\theta \max_\phi \sum_{n=1}^{N} \Big( \mathbb{E}_{d \sim p_{\text{true}}(d|q_n, r)} [\log D(d|q_n)] + \quad (1)$$

$$\mathbb{E}_{d \sim p_\theta(d|q_n, r)} [\log(1 - D(d|q_n))] \Big),$$

where the generative retrieval model $G$ is written as $p_\theta(d|q_n, r)$, directly and the discriminative retrieval $D$ estimates the probability of document $d$ being relevant to query $q$, which is given by the sigmoid function of the discriminator score

$$D(d|q) = \sigma(f_\phi(d, q)) = \frac{\exp(f_\phi(d, q))}{1 + \exp(f_\phi(d, q))} . \quad (2)$$

Let us leave the specific parametrisation of $f_\phi(d, q)$ to the next section when we discuss three specific IR tasks. From Eq. (1), we can see that the optimal parameters of the generative retrieval model and the discriminative retrieval model can be learned iteratively by maximising and minimising the same objective function, respectively.

*2.1.2 Optimising Discriminative Retrieval.* The objective for the discriminator is to *maximise* the log-likelihood of correctly distinguishing the true and generated relevant documents. With the observed relevant documents, and the ones sampled from the current optimal generative model $p_{\theta^*}(d|q,r)$, one can then obtain the optimal parameters for the discriminative retrieval model:

$$\phi^* = \arg\max_\phi \sum_{n=1}^{N} \Big( \mathbb{E}_{d \sim p_{\text{true}}(d|q_n, r)} \big[\log(\sigma(f_\phi(d, q_n))\big] +$$

$$\mathbb{E}_{d \sim p_{\theta^*}(d|q_n, r)} \big[\log(1 - \sigma(f_\phi(d, q_n)))\big] \Big), (3)$$

where if the function $f_\phi$ is differentiable with respect to $\phi$, the above is solved typically by stochastic gradient descent.

*2.1.3 Optimising Generative Retrieval.* By contrast, the generative retrieval model $p_\theta(d|q,r)$ intends to *minimise* the objective; it fits the underlying relevance distribution over documents $p_{\text{true}}(d|q,r)$ and based on that, randomly samples documents from the whole document set in order to *fool* the discriminative retrieval model.

It is worth mentioning that unlike GAN [13, 18], we design the generative model to directly generate known documents (in the document identifier space) not their features, because our work here intends to select relevant documents from a given document pool. Note that it is feasible to generate new documents (features, such as the value of BM25) by IRGAN, but to stay focused, we leave it for future investigation.

Specifically, while keeping the discriminator $f_\phi(q, d)$ fixed after its maximisation in Eq. (1), we learn the generative model via performing its minimisation:

$$\theta^* = \arg\min_\theta \sum_{n=1}^{N} \Big( \mathbb{E}_{d \sim p_{\text{true}}(d|q_n, r)} \left[ \log \sigma(f_\phi(d, q_n)) \right] +$$
$$\mathbb{E}_{d \sim p_\theta(d|q_n, r)} \left[ \log(1 - \sigma(f_\phi(d, q_n))) \right] \Big)$$
$$= \arg\max_\theta \sum_{n=1}^{N} \underbrace{\mathbb{E}_{d \sim p_\theta(d|q_n, r)} \left[ \log(1 + \exp(f_\phi(d, q_n))) \right]}_{\text{denoted as } J^G(q_n)}, \quad (4)$$

where for each query $q_n$ we denote the objective function of the generator as $J^G(q_n)$[1].

As the sampling of $d$ is discrete, it cannot be directly optimised by gradient descent as in the original GAN formulation. A common approach is to use policy gradient based reinforcement learning (REINFORCE) [42, 44]. Its gradient is derived as follows:

$$\nabla_\theta J^G(q_n)$$
$$= \nabla_\theta \mathbb{E}_{d \sim p_\theta(d|q_n, r)} \left[ \log(1 + \exp(f_\phi(d, q_n))) \right]$$
$$= \sum_{i=1}^{M} \nabla_\theta p_\theta(d_i|q_n, r) \log(1 + \exp(f_\phi(d_i, q_n)))$$
$$= \sum_{i=1}^{M} p_\theta(d_i|q_n, r) \nabla_\theta \log p_\theta(d_i|q_n, r) \log(1 + \exp(f_\phi(d_i, q_n)))$$
$$= \mathbb{E}_{d \sim p_\theta(d|q_n, r)} \left[ \nabla_\theta \log p_\theta(d|q_n, r) \log(1 + \exp(f_\phi(d, q_n))) \right]$$
$$\simeq \frac{1}{K} \sum_{k=1}^{K} \nabla_\theta \log p_\theta(d_k|q_n, r) \log(1 + \exp(f_\phi(d_k, q_n))), \quad (5)$$

where we perform a sampling approximation in the last step in which $d_k$ is the $k$-th document sampled from the current version of generator $p_\theta(d|q_n, r)$. With reinforcement learning terminology, the term $\log(1 + \exp(f_\phi(d, q_n)))$ acts as the reward for the policy $p_\theta(d|q_n, r)$ taking an action $d$ in the environment $q_n$ [38].

In order to reduce variance during the REINFORCE learning, we also replace the reward term $\log(1 + \exp(f_\phi(d, q_n)))$ by its advantage function:

$$\log(1 + \exp(f_\phi(d, q_n))) - \mathbb{E}_{d \sim p_\theta(d|q_n, r)} \left[ \log(1 + \exp(f_\phi(d, q_n))) \right],$$

where the term $\mathbb{E}_{d \sim p_\theta(d|q_n, r)} \left[ \log(1 + \exp(f_\phi(d, q_n))) \right]$ acts as the baseline function in policy gradient [38].

The overall logic of our proposed IRGAN solution is summarised in Algorithm 1. Before the adversarial training, the generator and discriminator can be initialised by their conventional models. Then during the adversarial training stage, the generator and discriminator are trained alternatively via Eqs. (22) and (3).

---
[1] Following [13], $\mathbb{E}_{d \sim p_\theta(d|q_n, r)}[\log(\sigma(f_\phi(d, q_n)))]$ is normally used instead for maximisation, which keeps the same fixed point but provides more sufficient gradient for the generative model.

**Algorithm 1** Minimax Game for IR (a.k.a IRGAN)

**Input:** generator $p_\theta(d|q, r)$; discriminator $f_\phi(x_i^q)$;
   training dataset $\mathcal{S} = \{x\}$
1: Initialise $p_\theta(d|q, r), f_\phi(q, d)$ with random weights $\theta, \phi$.
2: Pre-train $p_\theta(d|q, r), f_\phi(q, d)$ using $\mathcal{S}$
3: **repeat**
4:   **for** g-steps **do**
5:     $p_\theta(d|q, r)$ generates $K$ documents for each query $q$
6:     Update generator parameters via policy gradient Eq. (22)
7:   **end for**
8:   **for** d-steps **do**
9:     Use current $p_\theta(d|q, r)$ to generate negative examples and combine with given positive examples $\mathcal{S}$
10:    Train discriminator $f_\phi(q, d)$ by Eq. (3)
11:  **end for**
12: **until** IRGAN converges

## 2.2 Extension to Pairwise Case

In many IR problems, it is common that the labelled training data available for learning to rank are not a set of relevant documents but a set of ordered document pairs for each query, as it is often easier to capture users' relative preference judgements on a pair of documents than their absolute relevance judgements on individual documents (e.g., from a search engine's click-through log) [19]. Furthermore, if we use graded relevance scales (indicating a varying degree of match between each document and the corresponding query) rather than binary relevance, the training data could also be represented naturally as ordered document pairs.

Here we show that our proposed IRGAN framework would also work in such a pairwise setting for learning to rank. For each query $q_n$, we have a set of labelled document pairs $R_n = \{\langle d_i, d_j\rangle | d_i > d_j\}$ where $d_i > d_j$ means that $d_i$ is more relevant to $q_n$ than $d_j$. As in Section 2.1, we let $p_\theta(d|q, r)$ and $f_\phi(q, d)$ denote the generative retrieval model and the discriminative retrieval model respectively.

The generator $G$ would try to generate document pairs that are similar to those in $R_n$, i.e., with the correct ranking. The discriminator $D$ would try to distinguish such generated document pairs from those real document pairs. The probability that a document pair $\langle d_u, d_v\rangle$ being correctly ranked can be estimated by the discriminative retrieval model through a sigmoid function:

$$D(\langle d_u, d_v\rangle|q) = \sigma(f_\phi(d_u, q) - f_\phi(d_v, q))$$
$$= \frac{\exp(f_\phi(d_u, q) - f_\phi(d_v, q))}{1 + \exp(f_\phi(d_u, q) - f_\phi(d_v, q))} = \frac{1}{1 + \exp(-z)}, \quad (6)$$

where $z = f_\phi(d_u, q) - f_\phi(d_v, q)$. Note that $-\log D(\langle d_u, d_v\rangle|q) = \log(1 + \exp(-z))$ is exactly the pairwise ranking loss function used by the learning to rank algorithm RankNet [3]. In addition to the logistic function $\log(1 + \exp(-z))$, it is possible to make use of other pairwise ranking loss functions [7], such as the hinge function $(1 - z)_+$ (as used in Ranking SVM [16]) and the exponential function $\exp(-z)$ (as used in RankBoost [11]), to define the probability $D(\langle d_u, d_v\rangle|q)$.

If we use the standard cross entropy cost for this binary classifier as before, we have the following minimax game:

$$J^{G^*, D^*} = \min_\theta \max_\phi \sum_{n=1}^{N} \Big( \mathbb{E}_{o \sim p_{\text{true}}(o|q_n)} \left[ \log D(o|q_n) \right] + \quad (7)$$
$$\mathbb{E}_{o' \sim p_\theta(o'|q_n)} \left[ \log(1 - D(o'|q_n)) \right] \Big),$$

where $\mathbf{o} = \langle d_u, d_v \rangle$ and $\mathbf{o}' = \langle d'_u, d'_v \rangle$ are true and generated document pairs for query $q_n$ respectively.

In practice, to generate a document pair through generator $G$, we first pick a document pair $\langle d_i, d_j \rangle$ from $R_n$, take the lower ranked document $d_j$, and then pair it with a document $d_k$ selected from the unlabelled data to make a new document pair $\langle d_k, d_j \rangle$. The underlying rationale is that we are more interested in identifying the documents similar to higher ranked document $d_i$ as such documents are more likely to be relevant to the query $q_n$. The selection of the document $d_k$ is based on the criterion that $d_k$ should be more relevant than $d_j$ according to the current generative model $p_\theta(d|q, r)$. In other words, we would like to select $d_k$ from the whole document set to generate a document pair $\langle d_k, d_j \rangle$ which can imitate the document pair $\langle d_i, d_j \rangle \in R_n$.

Suppose that the generative model $p_\theta(d|q, r)$ is given by a softmax function (which is indeed used throughout Section 3, as we shall see later)

$$p_\theta(d_k|q, r) = \frac{\exp(g_\theta(q, d_k))}{\sum_d \exp(g_\theta(q, d))} , \quad (8)$$

where $g_\theta(q, d)$ is a task-specific real-valued function reflecting the chance of $d$ being generated from $q$. The probability of choosing a particular document $d_k$ could then be given by another softmax function:

$$G(\langle d_k, d_j \rangle | q) = p_\theta(\mathbf{o}'|q) = \frac{\exp\left(g_\theta(d_k, q) - g_\theta(d_j, q)\right)}{\sum_d \exp\left(g_\theta(d, q) - g_\theta(d_j, q)\right)}$$
$$= \frac{\exp(g_\theta(d_k, q))}{\sum_d \exp(g_\theta(d, q))} = p_\theta(d_k|q, r) . \quad (9)$$

In this special case, $G(\langle d_k, d_j \rangle | q)$ happens to be equal to $p_\theta(d_k|q, r)$, which is simple and reasonable. In general, the calculation of $G(\langle d_k, d_j \rangle | q)$ probably involves both $p_\theta(d_k|q, r)$ and $p_\theta(d_j|q, r)$. For example, one alternative way is to sample $d_k$ only from the documents more relevant to the query than $d_j$, and let $G(\langle d_k, d_j \rangle | q)$ be directly proportional to $\max(p_\theta(d_k|q, r) - p_\theta(d_j|q, r), 0)$.

This generative model $p_\theta(d|q, r)$ could be trained by the REINFORCE algorithm [42, 44] in the same fashion as we have explained in Section 2.1.

## 2.3 Discussion

It can be proved that when we know the true relevance distribution exactly, the above minimax game of IRGAN, both pointwise and pairwise, has a Nash equilibrium in which the generator perfectly fits the distribution of true relevant documents (i.e., $p_\theta(d|q, r) = p_{\text{true}}(d|q, r)$ in the pointwise case and $p_\theta(\mathbf{o}'|q) = p_{\text{true}}(\mathbf{o}|q)$ in the pairwise case), while the discriminator cannot distinguish generated relevant documents from the true ones (i.e., the probability of $d$ being relevant to $q$, $D(d|q)$ in the pointwise case or $D(\mathbf{o}'|q)$ in the pairwise case, is always $\frac{1}{2}$) [13]. However, in practice, the true distribution of relevant documents is unknown, and in such a situation, how the generative/discriminative retrieval models converge to achieve such an equilibrium is still an open problem in the current research literature [13, 14]. In our empirical study of IRGAN, we have found that depending on the specific task, the generative and discriminative retrieval models may reach different levels of performance; and at least one of them would be significantly improved in comparison to the corresponding original model without adversarial training.

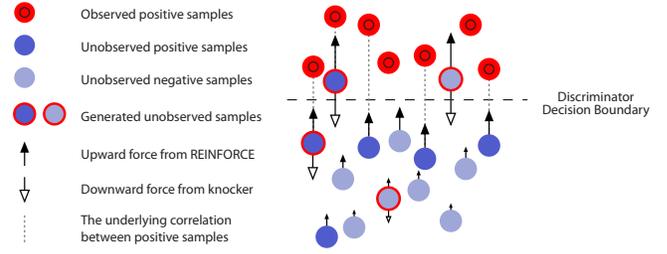

Figure 1: An illustration of IRGAN training.

How do the discriminator and the generator help each other? For the positive documents, observed or not, their relevance scores given by the discriminator $f_\phi(q, d)$ and the conditional probabilistic density $p_\theta(d|q, r)$ are likely to be somewhat positively correlated. In each epoch of training, the generator tries to generate samples close to the discriminator's decision boundary to confuse its training next round, while the discriminator tries to score down the generated samples. Since there exists positive correlations between the positive but unobserved (i.e., the true-positive) samples and (part of) the observed positive samples, the generator should be able to learn to push upwards these positive but unobserved samples faster than other samples with the signal from the discriminator.

To understand this process further, let us draw an analogy with a knocker kicking the floating soap in the water, as illustrated in Figure 1. There exist linking lines (i.e. positive correlations) between the unobserved positive soaps to the observed positive soaps that keep floating on the water surface (i.e. decision boundary of the discriminator) permanently. The discriminator acts as the knocker that kicks down the floating-up soaps, while the generator acts as the water that *selectively* floats the soaps up to the water surface. Even if the generator cannot perfectly fit the conditional data distribution, there could be still a dynamic equilibrium, which is obtained when the distribution of the positive and negative unobserved soaps get stable at different depth of the water. Since the unobserved positive soaps are linked to those observed positive soaps staying on the water surface, overall they should be able to reach higher positions than the (unobserved) negative soaps in the end.

Just like other GANs [12, 13, 44], the complexity of IRGAN training highly depends on the number of GAN iterations, each of which is of linear complexity $O(NKM)$ with respect to the number of candidate documents $M$. Such a complexity can largely be reduced to $O(NK \log M)$ by applying hierarchical softmax [28] in the sampling process of the generator.

## 2.4 Links to Existing Work

Let us continue our discussion on related work started in Section 1 and make comparisons with existing techniques in a greater scope.

*2.4.1 Generative Adversarial Nets.* Generative Adversarial Nets [13] were originally proposed to generate continuous data such as images. Our work is different in the following three aspects. First, the generative retrieval process is stochastic sampling over discrete data, i.e., the candidate documents, which is different from the deterministic generation based on the sampled noise signal in the original GAN. Specifically, as shown in Eq. (20), for each query $q_n$, the objective of the generative retrieval model is to minimise the expectation of the reward signal from the discriminative retrieval over the generated document distribution, while in the original GAN, the reward signal is solely dependent on a single generated instance. Second, our learning process of the generative retrieval model is based on the REINFORCE algorithm, a stochastic policy gradient technique in the field of reinforcement learning [44]. In IRGAN, the generative retrieval model can be regarded as an actor which takes an action of selecting a candidate document in a given environment of the query; the discriminative retrieval model can be regarded as a critic which performs a judgement whether the query-document pair is relevant enough. Third, during training, the conflict between ground-truth documents and generated documents is quite common, because documents are discrete and the candidate set is finite, which departs from the continuous (infinite) space for images or the extremely huge discrete (nearly infinite) space for text sequences [44]. Fourth, we also propose a pairwise discriminative objective, which is unique for IR problems.

Our work is also related to conditional GAN [29] as our generative and discriminative models are both conditional on the query.

### 2.4.2 MLE based Retrieval Models.
For unsupervised learning problems that estimate the data p.d.f. $p(x)$ and supervised learning problems that estimate the conditional p.d.f. $p(y|x)$, maximum likelihood estimation (MLE) plays as the standard learning solution [30]. In IR, MLE is also widely used as an estimation method for many relevance features or retrieval models [1], such as Term Frequency (TF), Mixture Model (MM) [49], and Probabilistic Latent Semantic Indexing (PLSI) [17]. In this paper, we provide an alternative way of training and fusing retrieval models. First, the generative process is designed to fit the underlying true conditional distribution $p_{\text{true}}(d|q,r)$ via minimising the Jensen-Shannon divergence (as explained in [13]). Thus, it is natural to leverage GAN to distil a generative retrieval model to fit such an unknown conditional distribution using the observed user feedback data. Second, the unified training scheme of two schools of IR models offers the potential of getting better retrieval models, because (i) the generative retrieval adaptively provides different negative samples to the discriminative retrieval training, which is strategically diverse compared with static negative sampling [3, 34] or dynamic negative sampling using the discriminative retrieval model itself [4, 50, 51]; and (ii) the reward signal from the discriminative retrieval model provides strategic guidance on training the generative retrieval model, which is unavailable in traditional generative retrieval model training. From the generative retrieval's perspective, IRGAN is superior to traditional maximum likelihood estimation [18]. From the discriminative retrieval's perspective, IRGAN is able to exploit unlabelled data to achieve the effect of semi-supervised learning [36]. The advantages of employing two models working together have received more and more attention in recent research; one of the variations is *dual learning* [43] proposed for two-agent co-learning in machine translation etc.

It is also worth comparing IRGAN with pseudo relevance feedback [39, 45, 50], where the top retrieved documents are selected to refine the ranking result. The two techniques are quite different as (i) in pseudo relevance feedback the top retrieved documents are regarded as positive samples to train the ranker while in IRGAN the generator-picked documents are regarded as negative samples to train the ranker; (ii) in pseudo relevance feedback there is usually no further iterations while IRGAN involves many iterations of adversarial training.

### 2.4.3 Noise-Contrastive Estimation.
Our work is also related to noise-contrastive estimation (NCE) that aims to correctly distinguish the true data $(y, x) \sim p_{\text{data}}(y|x)$ from known noise samples $(y_n, x) \sim p_{\text{noise}}(y_n|x)$. NCE is proved to be equivalent with MLE when noise samples are abundant [15]. With finite noise samples for contrastive learning, NCE is usually leveraged as an efficient approximation to MLE when the latter is inefficient, for example when the p.d.f is built by large-scale softmax modelling.

Furthermore, self-contrastive estimation (SCE) [14], a special case of NCE when the noise is directly sampled from the current (or a very recent) version of the model. It is proved that the gradient of SCE matches that of MLE with no prerequisite of infinite noise samples, which is a very attractive property of SCE learning. Dynamic negative item sampling [46, 51] in top-N item recommendation with implicit feedback turns out to be a practical use case of SCE, with specific solution of efficient sampling strategies.

The emergence of GANs [13], including our proposed IRGAN, opens a door to learning generative and discriminative retrieval models simultaneously. Compared to NCE and SCE, the GAN paradigm enables two models to learn together in an adversarial fashion, i.e. the discriminator learns to distinguish the true samples from the generated (faked) ones while the generator learns to generate high-quality samples to fool the discriminator.

## 3 APPLICATIONS
In this section, we apply our IRGAN framework to three specific IR scenarios: (i) web search with learning to rank, (ii) item recommendation, and (iii) question answering.

As formulated in Section 2, the generator's conditional distribution $p_\theta(d_i|q,r) = \exp(g_\theta(q,d_i))/\sum_{d_j} \exp(g_\theta(q,d_j))$, i.e., Eq. (8), fully depends on the scoring function $g_\theta(q,d)$. In the sampling stage, the temperature parameter $\tau$ is incorporated in Eq. (8) as

$$p_\theta(d|q,r) = \frac{\exp(g_\theta(q,d)/\tau)}{\sum_{j \in I} \exp(g_\theta(q,d)/\tau)}, \quad (10)$$

where a lower temperature would make the sampling focus more on top-ranked documents. A special case is when the temperature is set to 0, which implies that the entropy of the generator is 0. In this situation, the generator simply ranks the documents in descending order and selects the top ones. More detailed study of $\tau$ will be given in Section 4.

The discriminator's ranking of documents, i.e., Eq. (2) for the pointwise setting and Eq. (6) for the pairwise setting, is fully determined by the scoring function $f_\phi(q,d)$.

The implementation of these two scoring functions, $g_\theta(q,d)$ and $f_\phi(q,d)$, are task-specific. Although there could be various implementations of $f_\phi(q,d)$ and $g_\theta(q,d)$ (e.g., $f_\phi(q,d)$ is implemented as a three-layer neural work while $g_\theta(q,d)$ is implemented as a factorisation machine [33]), to focus more on adversarial training, in this section we choose to implement them using the same function (with different sets of parameters)[2]:

$$g_\theta(q,d) = s_\theta(q,d) \quad \text{and} \quad f_\phi(q,d) = s_\phi(q,d), \quad (11)$$

and in the following subsections we will discuss the implementation of the relevance scoring function $s(q,d)$ for those three chosen IR scenarios.

---
[2] We will, however, conduct a dedicated experiment on the interplay between these two players using the scoring functions of different model complexity, in Section 4.1.

## 3.1 Web Search

Generally speaking, there are three types of loss functions designed for learning to rank in web search, namely, pointwise [31], pairwise [3] and listwise [6]. To our knowledge, the listwise approaches with a loss defined on document pairs and a list-aware weight added on document pairs, e.g., LambdaRank [5] and LambdaMART [4], often can achieve the best performance across various learning to rank tasks. Despite the variety of ranking loss functions, almost every learning to rank solution boils down to a scoring function $s(q, d)$.

In the web search scenario, each query-document pair $(q, d)$ can be represented by a vector $x_{q,d} \in \mathbb{R}^k$, where each dimension represents some statistical value of the query-document pair or either part of it, such as BM25, PageRank, TFIDF, language model score etc. We follow the work of RankNet [3] to implement a two-layer neural network for the score function:

$$s(q, d) = w_2^\top \tanh(W_1 x_{q,d} + b_1) + w_0 , \quad (12)$$

where $W_1 \in \mathbb{R}^{l \times k}$ is the fully-connected matrix for the first layer, $b_1 \in \mathbb{R}^l$ is the bias vector for the hidden layer, $w_2 \in \mathbb{R}^l$ and $w_0$ are the weights for the output layer.

## 3.2 Item Recommendation

Item recommendation is a popular data mining task that can be regarded as a generalised information retrieval problem, where the query is the user profile constructed from their past item consumption. One of the most important methodologies for recommender systems is collaborative filtering which explores underlying user-user or item-item similarity and based on which performs personalised recommendations [41]. In collaborative filtering, a widely adopted model is matrix factorisation [21], following which we define our scoring function for the preference of user $u$ (i.e. the query) to item $i$ (i.e. the document) as

$$s(u, i) = b_i + v_u^\top v_i , \quad (13)$$

where $b_i$ is the bias term for item $i$, $v_u, v_i \in \mathbb{R}^k$ are the latent vectors of user $u$ and item $i$ respectively defined in a $k$-dimensional continuous space. Here we omit the global bias and the user bias as they are reduced in the task of top-N item recommendation for each user[3].

To keep our discussion uncluttered, we have chosen a basic matrix factorisation model to implement, and it would be straightforward to replace it with more sophisticated models such as factorisation machines [33] or neural networks [8], whenever needed.

## 3.3 Question Answering

In question answering (QA) tasks [9], a question $q$ or an answer $a$ is represented as a sequence of words. Typical QA solutions aim to understand the natural language question first and then select/generate one or more answers which best match the question [9]. Among various QA tasks, the document-based QA task cab be regarded as a ranking process based on the matching score between two pieces of texts (for question and answer, respectively) [9]. Recently, end-to-end approaches to predicting the match of short text pairs have been proposed, by utilising neural networks, such as convolutional neural network (CNN) [9, 37] or long short-term memory neural network (LSTM) [40].

For any question-answer pair $(q, a)$, we can define a relevance score. Specifically, one can leverage a convolutional neural networks (CNN) to learn the representation of word sequences [20], where each word is embedded as a vector in $\mathbb{R}^k$. By aligning the word vectors, an $l$-word sentence can be considered as a matrix in $\mathbb{R}^{l \times k}$. Then, a representation vector of the current sentence is obtained through a max-pooling-over-time strategy after a convolution operation over the matrix of aligned embedding vectors, yielding $v_q$ and $v_a \in \mathbb{R}^z$, where $z$ is the number of convolutional kernels. The relevance score of such a question-answer pair can be defined as their cosine similarity, i.e.,

$$s(q, a) = \cos(v_q, v_a) = \frac{v_q^\top v_a}{|v_q| \cdot |v_a|} . \quad (14)$$

With the sentence representation and the scoring function defined above, the question answering problem is transformed into a query-document scoring problem in IR [37].

## 4 EXPERIMENTS

We have conducted our experiments[4] corresponding to the three real-world applications of our proposed IRGAN as discussed, i.e., web search, item recommendation, and question answering. As each of the three applications has its own background and baseline algorithms, this section about experiments is split into three self-contained subsections. We first test both the IRGAN-pointwise and IRGAN-pairwise formulations within a single task, web search; and then IRGAN-pointwise is further investigated in the item recommendation task where the rank bias is less critical, while IRGAN-pairwise is examined in the question answering task where the rank bias is more critical (usually only one answer is correct).

### 4.1 Web Search

*4.1.1 Experiment Setup.* Web search is an important problem in the IR field. Here we make use of the well-known benchmark dataset LETOR (LEarning TO Rank) [25] for webpage ranking to conduct our experiments.

Although standard learning to rank tasks assume explicit expert ratings for all training query-document pairs, implicit feedback from user interaction (such as the clicks information) is much more common in practical applications. This implies that we are usually faced with a relatively small amount of labelled data inferred from implicit feedback and a large amount of unlabelled data. In the unlabelled data, there could exist some hidden positive examples that have not been discovered yet. Thus, we choose to do experiments in a semi-supervised setting on the MQ2008-semi (Million Query track) collection in LETOR 4.0: other than the labelled data (judged query-document pairs), this collection also contains a large amount of unlabelled data (unjudged query-document pairs), which can be effectively exploited by our IRGAN framework.

Each query-document pair in the dataset is given a relevance level (−1, 0, 1 or 2). The higher the relevance level, the more relevant the query-document pair, except that −1 means "unknown". Each query-document pair is represented by a 46-dimensional vector of features (such as BM25 and LMIR). To evaluate our proposed IRGAN in the context of implicit feedback, we consider all the query-document pairs with relevance level higher than 0 as positive examples, and all the other query-document pairs (with relevance level −1 or 0) as unlabelled examples. According to our statistics, there are 784 unique queries in this dataset; on average each query is associated with about 5 positive documents and about 1,000 unlabelled documents. To construct the training and test sets, we perform a 4:1 random splitting. Both pointwise and pairwise IRGANs are evaluated based on this dataset.

---

[3]The user bias could be taken as a good baseline function for the advantage function in policy gradient (Eq. (22)) to reduce the learning volatility [38].

[4]The experiment code is provided at: https://github.com/geek-ai/irgan

Similar to RankNet [3], we adopt a neural network model with one hidden layer and tanh activation to learn the query-document matching score, where the size of the hidden layer equals to the size of features. Besides, both the generator and discriminator are trained from scratch.

In the experiments, we compare the generative retrieval model in our IRGAN framework with simple RankNet [3], LambdaRank [5], and the strong baseline LambdaMART [4] for which we use the RankLib[5] implementation. For the evaluation of those compared algorithms, we use standard ranking performance measures [7] such as Precision@N, Normalised Discounted Cumulative Gain (NDCG@N), Mean Average Precision (MAP) and Mean Reciprocal Ranking (MRR).

*4.1.2 Results and Discussions.* First, we provide the overall performance of all the compared learning to rank algorithms on the MQ2008-semi dataset in Table 1. In our IRGAN framework, we use the generative retrieval model to predict the distribution of the user preferred documents given a query and then carry out the ranking, which is identical to performing the softmax sampling with the temperature parameter set very close to 0. From the experimental results we can see clear performance improvements brought by our IRGAN approach on all the metrics.

Specifically, IRGAN-pairwise works better than IRGAN-pointwise on the metrics of Precision@3, NDCG@3 that focus on a few webpages at the very top of the ranked list, whereas IRGAN-pointwise performs better than IRGAN-pairwise on the metrics of Precision@10, NDCG@10 and MAP that take into account more webpages high in the ranked list. A possible explanation is that IRGAN-pointwise is targeted for the conditional distribution $p_{\text{true}}(d|q,r)$ which only concerns whether an individual document is relevant to the query, whereas IRGAN-pairwise cares about the whole ranking of the documents given the query.

It is worth mentioning that the dataset studied in our experiments comes with implicit feedback, which is common in real life applications including web search and online advertising. Traditional learning to rank methods like LambdaMART are not particularly effective in this type of semi-supervised setting, which may be due to its reliance on the ΔNDCG scoring for each document pair [5].

Moreover, since adversarial training is widely regarded as an effective but unstable technique, we further investigate the learning trend of our proposed approach. Figures 2 and 3 show the typical learning curves of the generative/discriminative retrieval models in IRGAN-pointwise and IRGAN-pairwise respectively. Here we only show the performance measured by Precision@5 and NDCG@5 for discussion; the other metrics exhibit a similar trend. We can observe that after about 150 epoches for IRGAN-pointwise and 60 epoches for IRGAN-pairwise of adversarial training, both Precision@5 and

---

[5]https://sourceforge.net/p/lemur/wiki/RankLib/

**Table 1: Webpage ranking performance comparison on the MQ2008-semi dataset, where ∗ means a significant improvement according to the Wilcoxon signed-rank test.**

|  | P@3 | P@5 | P@10 | MAP |
|---|---|---|---|---|
| MLE | 0.1556 | 0.1295 | 0.1029 | 0.1604 |
| RankNet [3] | 0.1619 | 0.1219 | 0.1010 | 0.1517 |
| LambdaRank [5] | 0.1651 | 0.1352 | 0.1076 | 0.1658 |
| LambdaMART [4] | 0.1368 | 0.1026 | 0.0846 | 0.1288 |
| IRGAN-pointwise | 0.1714 | 0.1657 | **0.1257** | **0.1915** |
| IRGAN-pairwise | **0.2000** | **0.1676** | 0.1248 | 0.1816 |
| Impv-pointwise | 3.82% | 22.56%* | 16.82%* | 15.50%* |
| Impv-pairwise | 21.14%* | 23.96%* | 15.98% | 9.53% |
|  | NDCG@3 | NDCG@5 | NDCG@10 | MRR |
| MLE | 0.1893 | 0.1854 | 0.2054 | 0.3194 |
| RankNet [3] | 0.1801 | 0.1709 | 0.1943 | 0.3062 |
| LambdaRank [5] | 0.1926 | 0.1920 | 0.2093 | 0.3242 |
| LambdaMART [4] | 0.1573 | 0.1456 | 0.1627 | 0.2696 |
| IRGAN-pointwise | 0.2065 | **0.2225** | **0.2483** | **0.3508** |
| IRGAN-pairwise | **0.2148** | 0.2154 | 0.2380 | 0.3322 |
| Impv-pointwise | 7.22% | 15.89% | 18.63% | 8.20% |
| Impv-pairwise | 11.53% | 12.19% | 13.71% | 2.47% |

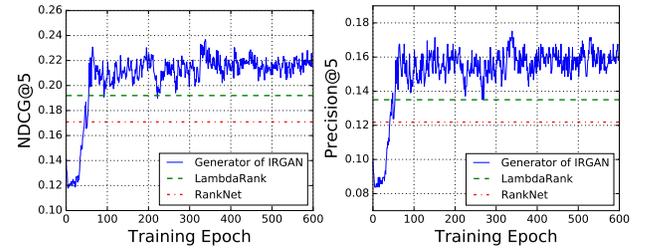

**Figure 2: Learning curves of the pointwise IRGAN on the web search task.**

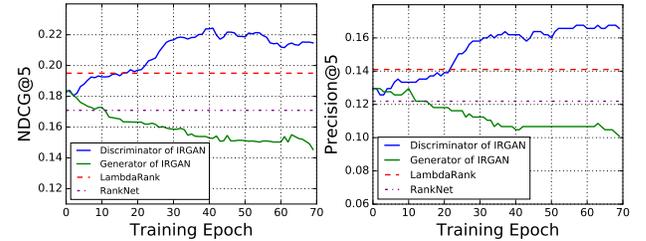

**Figure 3: Learning curves of the pairwise IRGAN on the web search task.**

NDCG@5 converge and the winner player consistently outperforms the best baseline LambdaRank.

Figure 4 shows how the ranking performance varies over the temperature parameter in Eq. (10) used by the generative retrieval model to sample negative query-document pairs for the discriminative retrieval model. We find the empirically optimal sampling temperature to be 0.2. The ranking performance increases when the temperature is tuned from 0 to the optimal value and then drops down afterwards, which indicates that *properly* increasing the aggressiveness (i.e. the tendency to focus on the top-ranked documents) of the generative retrieval model is important.

Furthermore, we study the impact of the model complexity of $f_\phi(q,d)$ and $g_\theta(q,d)$ upon the interplay between them. In Figure 5

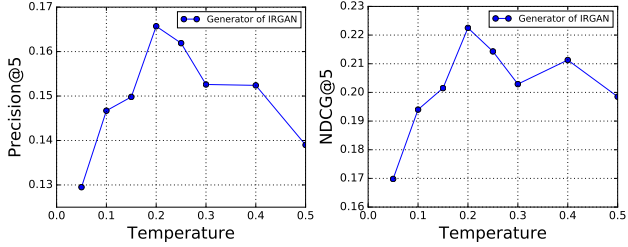

**Figure 4: Ranking performance with different sampling temperatures of pointwise IRGAN on the web search task.**

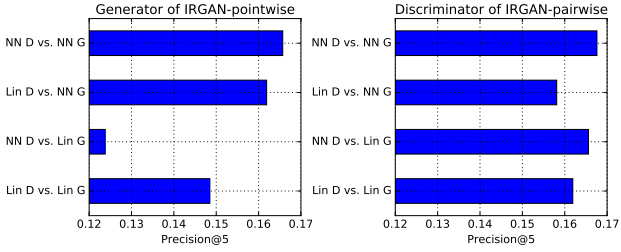

**Figure 5: Ranking performance for IRGAN with different generator and discriminator scoring functions.**

**Table 2: Characteristics of the datasets.**

| Dataset | Users | Items | Ratings |
|---|---|---|---|
| Movielens | 943 | 1,683 | 100,000 |
| Netflix | 480,189 | 17,770 | 100,480,507 |

we have compared different combinations of generative and discriminative model implementations (i.e., linear model and two-layer NN) under IRGAN-pointwise and IRGAN-pairwise, respectively. We observe that (i) for IRGAN-pointwise, the NN implemented generator works better than its linear version, while the NN implemented discriminator may not offer a good guidance if the generator has lower model complexity (i.e. linear); (ii) for IRGAN-pairwise, the NN implemented discriminator outperforms its linear version. This suggests that the model used for making the prediction (the generator in IRGAN-pointwise or the discriminator in IRGAN-pairwise) should be implemented with a capacity not lower than its opponent.

### 4.2 Item Recommendation

*4.2.1 Experiment Setup.* We conduct our experiments on two widely used collaborative filtering datasets: Movielens (100k) and Netflix. Their details are shown in Table 2. Following the experimental setting of [51], we regard the 5-star ratings in both Movielens and Netflix as positive feedback and treat all other entries as unknown feedback, because we mainly focus on the implicit feedbacks problem. For training and test data splitting, we apply a 4:1 random splitting on both datasets as in [51]. The factor numbers for matrix factorisation are 5 and 16 for Movielens and Netflix respectively.

Specifically, to help train the discriminative retrieval model, the generative retrieval model is leveraged to sample negative items (in the same number of positive items) for each user via Eq. (10) with the temperature parameter set to 0.2, which to some extent pushes the item sampling to the top ones. Then the training of the discriminative retrieval model is dictated by Eq. (3). On the other side of the game, the training of the generative retrieval model is performed by REINFORCE as in Eq. (22), which is normally implemented by the policy gradient on the sampled $K$ items from $p_\theta(d|q_n, r)$. In such a case, if the item set size is huge (e.g., more than $10^4$) compared with $K$, it is more practical to leverage importance sampling to force the generative retrieval model to sample (some) positive examples $d \in R_n$, so that the positive reward can be observed from REINFORCE and the generative retrieval model can be learned properly.

In the experiments, we compare IRGAN with Bayesian Personalised Ranking (BPR) [34] and a state-of-the-art LambdaRank based collaborative filtering (LambdaFM) [46] for top-N item recommendation tasks [46, 51]. Similar to the web search task, the performance measures are Precision@N, NDCG@N, MAP and MRR.

*4.2.2 Results and Discussion.* First, the overall performance of the compared approaches on the two datasets is shown in Tables 3

**Table 3: Item recommendation results (Movielens).**

|  | P@3 | P@5 | P@10 | MAP |
|---|---|---|---|---|
| MLE | 0.3369 | 0.3013 | 0.2559 | 0.2005 |
| BPR [34] | 0.3289 | 0.3044 | 0.2656 | 0.2009 |
| LambdaFM [46] | 0.3845 | 0.3474 | 0.2967 | 0.2222 |
| IRGAN-pointwise | 0.4072 | 0.3750 | 0.3140 | 0.2418 |
| Impv-pointwise | 5.90%* | 7.94%* | 5.83%* | 8.82%* |
|  | NDCG@3 | NDCG@5 | NDCG@10 | MRR |
| MLE | 0.3461 | 0.3236 | 0.3017 | 0.5264 |
| BPR [34] | 0.3410 | 0.3245 | 0.3076 | 0.5290 |
| LambdaFM [46] | 0.3986 | 0.3749 | 0.3518 | 0.5797 |
| IRGAN-pointwise | 0.4222 | 0.4009 | 0.3723 | 0.6082 |
| Impv-pointwise | 5.92%* | 6.94%* | 5.83%* | 4.92%* |

**Table 4: Item recommendation results (Netflix).**

|  | P@3 | P@5 | P@10 | MAP |
|---|---|---|---|---|
| MLE | 0.2941 | 0.2945 | 0.2777 | 0.0957 |
| BPR [34] | 0.3040 | 0.2933 | 0.2774 | 0.0935 |
| LambdaFM [46] | 0.3901 | 0.3790 | 0.3489 | 0.1672 |
| IRGAN-pointwise | 0.4456 | 0.4335 | 0.3923 | 0.1720 |
| Impv-pointwise | 14.23%* | 14.38%* | 12.44%* | 2.87%* |
|  | NDCG@3 | NDCG@5 | NDCG@10 | MRR |
| MLE | 0.3032 | 0.3011 | 0.2878 | 0.5085 |
| BPR [34] | 0.3077 | 0.2993 | 0.2866 | 0.5040 |
| LambdaFM [46] | 0.3942 | 0.3854 | 0.3624 | 0.5857 |
| IRGAN-pointwise | 0.4498 | 0.4404 | 0.4097 | 0.6371 |
| Impv-pointwise | 14.10%* | 14.27%* | 13.05%* | 8.78%* |

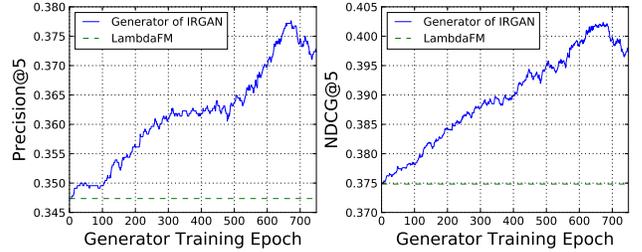

**Figure 6: Learning curve of precision and NDCG of the generative retrieval model for the top-5 item recommendation task on the Movielens dataset.**

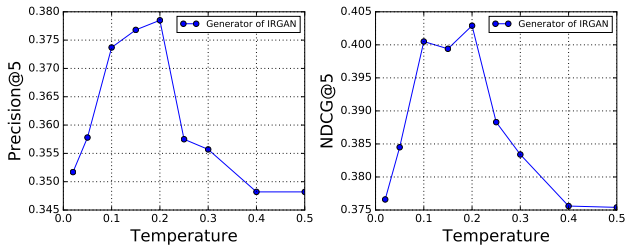

**Figure 7: Ranking performance with different sampling temperatures on the Movielens dataset.**

and 4. From the experimental results, we can observe that IRGAN achieves statistically significant improvements across all the evaluation metrics and all the datasets. Note that the generative retrieval model in IRGAN does not explicitly learn to optimise the final ranking measures like what LambdaFM does, it still performs consistently better than LambdaFM. Our explanation is that the adversarial training provides both models a higher learning flexibility than the single-model training of LambdaFM or BPR.

We further investigate the learning trend of the proposed approach. The learning curves are shown in Figure 6 for Precision@5 and NDCG@5. The experimental results demonstrate a reliable training process where IRGAN owns a consistent superiority over the baseline LambdaFM from the beginning of adversarial training. As for this case the curves are not as stable as those in web search (Figure 3), one can adopt the early stopping strategy based on a validation set.

In addition, as shown in Figure 7, we also investigate how the performance varies w.r.t. the sampling temperature in Eq. (10), which is consistent with our observations in the web search task.

### 4.3 Question Answering

*4.3.1 Experiment Setup.* InsuranceQA [10] is one of the most studied question-answering dataset. Its questions are submitted from real users and the high-quality answers are composed by professionals with good domain knowledge. So the candidate answers are usually randomly sampled from the whole answers pool (whereas other QA datasets may have a small-size fixed candidate answers for each single question). Thus InsuranceQA is suitable for testing our sampling/generating strategy. There are a training set, a development set, and two test sets (test-1 and test-2) in the published corpus. 12,887 questions are included in the training set with correct answers, while the development set have 1,000 unseen question-answer pairs and the two test sets consist of 1,800 pairs. The system is expected to find the *one and only* real answer from 500 candidate answers under the Precision@1 metric. As we have found from the web search task that IRGAN-pairwise works better for top-ranked documents, we concentrate on the former in the QA task experiments.

To focus on evaluating the effectiveness of IRGAN, we use a simple convolutional layer on the basic embedding matrix of a question sentence or an answer sentence. A representation vector of the current sentence is distilled from a max-pooling strategy after convolution [20], yielding $\boldsymbol{v}_q$ and $\boldsymbol{v}_a$ in Eq. (14). The matching probability of such a question-answer pair is given by the cosine distance, which is similar to the basic QA-CNN model [9].

In detail, the embedding of each word is initialised as a 100-dimension random vector. In the convolutional layer, the window size of the convolution kernel is set to (1, 2, 3, 5). After the convolutional layer, the max-pooling-over-time strategy is adopted [20], where each feature map will be pooled as a scalar since its convolution kernel width is the same as the embedding vector. The performance on the test set is calculated by the model in the epoch with the best performance evaluated on the development set. Our IRGAN solution would load pre-trained models as the initial parameters for both generator and discriminator. A question with multiple answers are considered as multiple questions each with a single corresponding answer, which means that for each question-answer pair only the feeding positive answer is observed by the current discriminator but the other positive answers are not.

**Table 5: The Precision@1 of InsuranceQA.**

|  | test-1 | test-2 |
|---|---|---|
| QA-CNN [9] | 0.6133 | 0.5689 |
| LambdaCNN [9, 51] | 0.6294 | 0.6006 |
| IRGAN-pairwise | 0.6444 | 0.6111 |
| Impv-pairwise | 2.38%* | 1.75% |

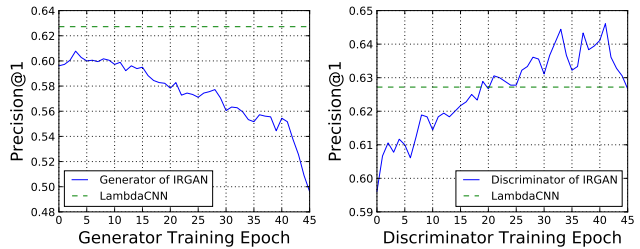

**Figure 8: The experimental results in QA task.**

*4.3.2 Results and Discussion.* As shown in Table 5, IRGAN outperforms both the basic CNN model with a random sampling strategy (QA-CNN) and the enhanced CNN model with a dynamic negative sampling strategy (LambdaCNN) [9, 51]. The learning curves of the two models are shown in Figure 8, which is evaluated on the test-1 set. The performance of the discriminative retrieval model in IRGAN-pairwise is better than LambdaCNN while the generative retrieval model tends to perform less effectively during the pairwise adversarial training. A reason for the worse generator could be the sparsity of the answers distribution, i.e., each question usually has only one correct answer and many more weak negative answers. Due to such a sparsity, the generator may fail to get a positive feedback from the discriminator. An inspection of the sampled answers from LambdaCNN and IRGAN has revealed that about 1/3 of their samples are different. This suggests the effectiveness of independently modelling the negative generator.

## 5 CONCLUSIONS

In this paper, we have proposed the IRGAN framework that unifies two schools of information retrieval methodologies, i.e., generative models and discriminative models, via adversarial training in a minimax game. Such an adversarial training framework takes advantages from both schools of methodologies: (i) the generative retrieval model is guided by the signals obtained from the discriminative retrieval model, which makes it more favourable than the non-learning methods or the maximum likelihood estimation scheme; (ii) the discriminative retrieval model could be enhanced to rank top documents better via strategic negative sampling from the generator. Overall, IRGAN provides a more flexible and principled training environment that combines these two kinds of retrieval models. Extensive experiments were conducted on four real-world datasets in three typical IR tasks, namely web search, item recommendation, and question answering. Significant performance gains were observed in each set of experiments.

Despite the great empirical success of GAN [13], there are still many questions with regard to its theoretical foundation remaining to be answered by the research community. For example, it is "not entirely clear" why GAN can generate sharper realistic images than alternative techniques [12]. Our exploration of adversarial training for information retrieval in the proposed IRGAN framework has suggested that different equilibria could be reached in the end depending on the task and setting. In the pointwise version of IRGAN, the generative retrieval model gets improved more than the discriminative retrieval model, but we have an opposite observation in the pairwise case. This phenomenon certainly warrants further investigation.

For future work, further experiments on more real-world datasets will be conducted. We also plan to extend our framework and test it over the generation of word tokens. One possible direction is to delve into the word weighting schemes such as [32, 35, 48] learned from the IRGAN generative retrieval model and then derive new ranking features on that basis. Furthermore, the language models could be re-defined along with GAN training, where new useful word patterns might emerge.

**Acknowledgement** We thank Geek.AI for hosting this collaborative project. The work done by SJTU is financially supported by NSFC (61632017) and Shanghai Sailing Program (17YF1428200).

# APPENDIX

In the appendix, we start from the discussion of the original GAN [13] formulation for continuous data generation and then a review of our IRGAN for discrete data generation. Our analysis of the gradients of two GANs illutrates their key differences. We further present our derivations and clarify that the two different GAN settings would lead to different specific implementations of the objective function in practice.

## A GAN FOR CONTINUOUS DATA

In the original GAN [13], the objective is

$$J(\theta, \phi) = \mathbb{E}_{x \sim p_{\text{data}}(x)}[\log D_\phi(x)] + \mathbb{E}_{z \sim p_z(z)}[\log(1 - D_\phi(G_\theta(z)))], \quad (15)$$

where the $\theta$-parameterized generator $G_\theta$ and the $\phi$-parameterized discriminator $D_\phi$ play a minimax game. The generator is formulated as a deterministic neural net that maps a noise vector $z \sim p_z(z)$ to the data instance $x$, and the discriminator is formulated as a sigmoid-based binary classifier

$$D_\phi(x) = \sigma(f_\phi(x)) = \frac{1}{1 + \exp(-f_\phi(x))} \quad (16)$$

where $f_\phi(x)$ could be any scoring function to calculate the logit value of the sigmoid.

The discriminator is trained to maximize such an objective with the fixed generator $G_\theta$

$$\phi^* = \arg\max_\phi J(\theta, \phi) \quad (17)$$

while the generator is trained to minimize such an objective with the fixed discriminator $D_\phi$

$$\theta^* = \arg\min_\theta J(\theta, \phi) = \arg\min_\theta \mathbb{E}_{z \sim p_z(z)}[\log(1 - D_\phi(G_\theta(z)))]. \quad (18)$$

There is an ideal equilibrium for above minimax game setting, i.e., the generated data distribution equals to the true data distribution and the discriminator returns a 0.5 judge for each generated data instance. As claimed in the original GAN paper [13], the term $\log(1 - D_\phi(G_\theta(z)))$ is normally replaced by $-\log D_\phi(G_\theta(z))$ to provide more sufficient gradient for the generator in the early training stage[6]. With Eq. (16) the generator gradient is written as

$$\nabla_\theta J(\theta, \phi) = \mathbb{E}_{z \sim p_z(z)}\Big[ - \nabla_\theta \log D_\phi(G_\theta(z))\Big]$$
$$= \mathbb{E}_{z \sim p_z(z)}\Big[ - \nabla_\theta \log \sigma(f_\phi(G_\theta(z)))\Big]$$
$$= \mathbb{E}_{z \sim p_z(z)}\Big[ - \frac{1}{\sigma(f_\phi(G_\theta(z)))} \nabla_\theta \sigma(f_\phi(G_\theta(z)))\Big]$$
$$= \mathbb{E}_{z \sim p_z(z)}\Big[ - \frac{\sigma(f_\phi(G_\theta(z)))(1 - \sigma(f_\phi(G_\theta(z))))}{\sigma(f_\phi(G_\theta(z)))} \nabla_\theta f_\phi(G_\theta(z))\Big]$$
$$= \mathbb{E}_{z \sim p_z(z)}\Big[ - \big(1 - \sigma(f_\phi(G_\theta(z)))\big) \nabla_\theta f_\phi(G_\theta(z))\Big]$$
$$= \mathbb{E}_{z \sim p_z(z)}\Big[ - \big(1 - \sigma(f_\phi(G_\theta(z)))\big) \nabla_{G_\theta(z)} f_\phi(G_\theta(z)) \nabla_\theta G_\theta(z)\Big] \quad (19)$$

---
[6]Although later WGAN proves that after using $\log D_\phi(G_\theta(z))$, the new training objective is not consistent with the original GAN, this setting is still the most widely adopted so far. Thus our implementation is also based on this setting.

A key step for calculating the above gradient is to calculate the term $\nabla_{G_\theta(z)} f_\phi(G_\theta(z))$, i.e., the gradient of the discriminator's prediction with respect to the generated data itself. This could only work on continuous data such as image pixels and speech audios.

## B GAN FOR DISCRETE DATA AND IMPLEMENTATION DETAILS

Another type of GAN, aiming at generating discrete data, such as SeqGAN [44] and the one presented here for IR tasks, departs from the above gradient update. The key difference lies in that, in the proposed IRGAN, our generator is modeled as a reinforcement learning policy to select a candidate document $d$ at the state, given query $q_n$, and a relevance metric $r$, which is trained via policy gradients. To make our discussion self-contained, we re-produce the optimization formulas for the generator here from the main text:

$$\theta^* = \arg\min_\theta \sum_{n=1}^{N} \Big( \mathbb{E}_{d \sim p_{\text{true}}(d|q_n,r)} \big[\log \sigma(f_\phi(d, q_n))\big]$$
$$+ \mathbb{E}_{d \sim p_\theta(d|q_n,r)} \big[\log(1 - \sigma(f_\phi(d, q_n)))\big]\Big)$$
$$= \arg\min_\theta \sum_{n=1}^{N} \mathbb{E}_{d \sim p_\theta(d|q_n,r)}\big[\log(1 - \sigma(f_\phi(d, q_n)))\big], \quad (20)$$

where, as mentioned in Footnote 1, a replacing trick [13] is typically adopted in practice. If we follow this, the objective of generator turns to

$$\theta^* = \arg\max_\theta \sum_{n=1}^{N} \mathbb{E}_{d \sim p_\theta(d|q_n,r)}\big[\log \sigma(f_\phi(d, q_n))\big]. \quad (21)$$

Regarding $V(d, q_n) \equiv \log \sigma(f_\phi(d, q_n))$ as the action value function[7], the generator of discrete data is trained via REINFORCE algorithm

$$\nabla_\theta J^G(q_n) = \nabla_\theta \mathbb{E}_{d \sim p_\theta(d|q_n,r)}\big[\log \sigma(f_\phi(d, q_n))\big]$$
$$= \sum_{i=1}^{M} \nabla_\theta p_\theta(d_i|q_n, r) \log \sigma(f_\phi(d, q_n))$$
$$= \sum_{i=1}^{M} p_\theta(d_i|q_n, r) \nabla_\theta \log p_\theta(d_i|q_n, r) \log \sigma(f_\phi(d, q_n))$$
$$= \mathbb{E}_{d \sim p_\theta(d|q_n,r)}\big[\nabla_\theta \log p_\theta(d|q_n, r) \log \sigma(f_\phi(d, q_n))\big]$$
$$\simeq \frac{1}{K} \sum_{k=1}^{K} \nabla_\theta \log p_\theta(d_k|q_n, r) \log \sigma(f_\phi(d, q_n)). \quad (22)$$

**Understanding the Gradient** However Eq. (22) is problematic if adopting $\log \sigma(f_\phi(d, q_n))$ in our discrete case as the reward signal. The reason is that when the action value term is in a large magnitude, the REINFORCE training could be very unstable, resulting in parameter explosion. In fact, such a kind of parameter explosion is likely to occur in practice because at the beginning of GAN training, the discriminator tends to return a large negative logit $f_\phi(q_n, d)$ for the generator's selection $d$, resulting the sigmoid value $\sigma(f_\phi(d, q_n))$ to be close to 0, also reported in [13], i.e., a large negative logarithm value $\log \sigma(f_\phi(d, q_n))$.

---
[7]This is a one-step RL task, thus the reward (function) is the value (function).

The above issue, nonetheless, does not appear in the original GAN for continuous data. As in Eq. (19), when $\sigma(f_\phi(G_\theta(z)))$ is very close to 0, the term $1 - \sigma(f_\phi(G_\theta(z))) \to 1$ and the gradient works normally.

**IRGAN Reward Implementation** To address the above training issues in IRGAN, in our experimental implementation, we drop the logarithm operation in the discriminator to make the value function be

$$\hat{V}(d, q_n) = \sigma(f_\phi(d, q_n)) . \qquad (23)$$

Typically a baseline or advantage function is used to reduce the variance of policy gradient. Our final value function is implemented as

$$\tilde{V}(d, q_n) = 2 \cdot \sigma(f_\phi(d, q_n)) - 1 , \qquad (24)$$

where value 1 is a baseline. We obtain promising results as reported in our experiments consistently across three application cases. The implementation and sample code are given in the IRGAN Github repository. Note that our SeqGAN implementation also applies such a reward design (see Eq. (2) in [44]) to reliably generate sequence discrete data. Reward shaping is one of the important subjects in reinforcement learning[8]. For example, when training a simulated robot to run in an environment as long as possible, the most generic form of reward is 1 if the robot is running, 0 otherwise. But in practice, to ease the training, one need design a bunch of reward functions considering angles, speed, height etc. We leave more theoretical investigation on reward shaping of the general information retrivel and text generation problems for future work.

---

[8]Ng, Andrew Y., Daishi Harada, and Stuart Russell. Policy invariance under reward transformations: Theory and application to reward shaping. ICML 1999.